\documentclass[aps,prl,floatfix,twocolumn,showpacs]{revtex4}
\usepackage{graphicx} \usepackage{amsmath} \usepackage{amssymb}
\newcommand{\BEQ}{\begin{equation}}
\newcommand{\EEQ}{\end{equation}}
\newcommand{\BEA}{\begin{eqnarray}}
\newcommand{\EEA}{\end{eqnarray}}

\renewcommand{\d}{{\rm d}}
\renewcommand{\t }{ \tau }

\newcommand{\R}{\bar{R}}

\newcommand{\infi}{ {\scriptstyle{\infty}} }

\newcommand{\eps}{\varepsilon}

\newcommand{\mc}{\bar{c}}

\begin{document}
\draft
\title
{Employing feedback in adiabatic quantum dynamics}
\author{Armen E. Allahverdyan$^1$ and Guenter Mahler$^2$}
\affiliation{$^1$Yerevan Physics Institute,
Alikhanian Brothers Street 2, Yerevan 375036, Armenia,\\
$^2$Institute of Theoretical Physics I, University of Stuttgart, 
Pfaffenwaldring 57, 70550 Stuttgart, Germany}

\begin{abstract} We study quantum adiabatic dynamics, where the slowly
moving field is influenced by system's state (feedback). The information
for the feedback is gained from non-disturbating measurements done on an
ensemble of identical non-interacting systems.  The situation without
feedback is governed by the adiabatic theorem: adiabatic energy level
populations stay constant, while the adiabatic eigenvectors get a
specific phase contribution (Berry phase).  However, under feedback the
adiabatic theorem does not hold: the adiabatic populations satisfy a
closed equation of motion that coincides with the replicator dynamics
well-known by its applications in evolutionary game theory. The feedback
generates a new gauge-invariant adiabatic phase, which is free of the
constraints on the Berry phase (e.g., the new phase is non-zero even for
real adiabatic eigenfunctions).

\end{abstract}



\maketitle

The adiabatic theorem of quantum mechanics governs the evolution of a
quantum system subject to slowly varying external fields. Its
applications span a vast array of fields, such as two-level systems
(nuclei undergoing magnetic resonance or atoms interacting with a laser
field), quantum field theory (where a low-energy effective theory is
derived by integrating out fast, high-energy degrees of freedom), and
Berry's phase \cite{messiah,berry,sj_ajp}. This phase and
the adiabatic theorem also find applications in quantum information
processing \cite{geophase_computing,farhi}. For a recent discussion on
the validity of the adiabatic approach see \cite{adia}.

A general perspective of the quantum adiabatic physics is that it
studies a system subject to a slow, {\it open loop} (i.e., no feedback)
control, where the evolution of the external fields is given {\it a
priori} via time-dependent parameters of the system Hamiltonian. In view
of numerous application of this setup, it is natural to wonder about the
quantum adiabatic {\it closed-loop} control, where the external
controlling fields evolve under feedback from the controlled quantum
system. Any feedback needs information on the evolution of the system.
This information is to be gained via measurements, which in the quantum
situation are normally related with unpredictable disturbances and
irreversibility. Thus quantum control has been restricted to open-system
dynamics \cite{jacobs}.

However, also quantum measurements can be non-disturbing, if $N$
non-interacting quantum particles (spins, {\it etc}) couple to the
proper measuring apparatus.  For $N\gg 1$ [analog of the classical
limit] one can measure single-particle observables (almost) without
disturbing the single-particle density matrix, since the disturbance
caused by such measurements scales as $\frac{1}{N\sigma^2}$, where
$\sigma$ is the measurement precision \cite{poulin}.  The knowledge of
these observables allows to implement feedback \cite{weak}.
Non-disturbing measurements on ensembles of few-level systems are
routinely employed in NMR physics (e.g., in ensemble computation) and
quantum optics \cite{wo,jajones,broek}. 

Here we develop adiabatic theory under feedback obtained via such
non-disturbing measurements.

{\it Basic equations.} Consider a $d$-level quantum system described by
a pure state $|\psi\rangle$ (generalization to mixed states is indicated
below).  The system evolves according to the Schroedinger equation with
Hamiltonian $H[R(t)]$, where $R(t)$ is a classical controllling
parameter ($\hbar=1$):
\BEA
\label{1}
i|\dot{\psi}(t)\rangle=H[R(t)]\,|\psi(t)\rangle.
\EEA
By means of a continuous non-disturbing measurement performed on
an ensemble of identical, non-interacting systems (each one described by
$|\psi(t)\rangle$) one finds the average $\langle
\psi(t)|A|\psi(t)\rangle$ of a monitoring observable $A$
(in NMR physics $A$ typically corresponds to the
magnetization). This average enters the feedback dynamics of $R$
\BEA
\dot{R}=\eps F\left(R,\langle \psi(t)|A|\psi(t)\rangle\right),
\label{2}
\EEA
where $\eps\ll 1$ is a small dimensionless parameter. We assume that
$F(.,.)$ is bounded from above, which means that $R$ is a slow variable:
its derivative is bounded by a small number. For $F=F(R)$ (no feedback)
we recover the standard adiabatic setup.  The dynamics (\ref{1})
conserves the purity of $|\psi(t)\rangle$, but the overlap $\langle
\phi(t)|\psi(t)\rangle$ between two different wave-functions is not
conserved in time, since $H$ depends on $|\psi(t)\rangle\langle\psi(t)|$
via (\ref{2}). 

For a particular case $F=-\langle \psi|\partial_R H|\psi\rangle$, Eqs.~(\ref{1}, \ref{2}) can be
viewed as a {\it hybrid} dynamics, where a classical particle with
coordinate $R$ performs an overdamped motion and couples to the quantum
system.  Then $\eps$ in (\ref{2}) corresponds to an inverse damping
constant, while $\langle \psi|\partial_R H|\psi\rangle$ is the average
(mean-field) force acting on the particle.  Hybrid theories are
frequently employed in optics and chemical physics; see \cite{hall} for
the current state of art. 

Let us now introduce the adiabatic eigenresolution
of the Hamiltonian for a fixed value of $R$ ($n=1,..,d$):
\BEA
\label{3}
H[R]\, | n[R] \rangle = E_n[R]\, | n[R] \rangle, ~~~~~~
\langle n[R]\,|\,m[R]\rangle=\delta_{mn}.
\EEA
For simplicity we assume that the adiabatic energy levels are
not degenerate. The representation (\ref{3}) has a gauge-freedom:
$| n[R] \rangle\,\to\,e^{i\alpha_n[R]}\,
| n[R] \rangle$, where $\alpha_n[R]$ is an 
arbitrary single-valued function depending on $n$ and $R$.
All observables should be gauge-invariant. Expand $|\psi(t)\rangle$ as
\begin{gather}
\label{5}
|\psi(t)\rangle ={\sum}_n c_n(t) \, e^{i\gamma_n(t)}\, |n[R(t)]\rangle, \\
\gamma_n(t)\equiv -\int_0^t \d t'\, E_n[R(t')], \,\,\,
c_n(t) =\langle n[R(t)]\,|\psi\rangle\, e^{-i\gamma_n(t)},\nonumber
\end{gather}
where $\gamma_n(t)$ are the dynamical phases, while $c_n$ are the adiabatic amplitudes.
One gets from (\ref{1}, \ref{2}, \ref{5}):
\BEA
\label{6}
\dot{c}_n 
=-\eps{\sum}_k c_k\langle n| k'\rangle\,F(R,c,e^{i\Delta \gamma(t)})\, e^{i(\gamma_k(t)-\gamma_n(t))},
\label{7}
\EEA
where $|k'\rangle=\partial_R|k[R]\rangle$.  The amplitudes $c_n$ and $R$
are slow variables, since, e.g., $|\dot{c}_n|$ is bounded from above by
the small $\eps$ in (\ref{7}).  However, the contribution from the
dynamical phases $\gamma_n$ changes fast, since on the slow time
$\tau=\eps t$ it behaves as $\sim e^{i\tau/\eps}$; see (\ref{5}).  If
the spacings between the adiabatic energy levels $E_n[R]$ remain large
enough, the existence of some intermediate time $\tau_f$ is guaranteed,
over which the dynamical phase contribution performs many oscillations,
but $c_n$ and $R$ do not change appreciably.  The adiabatic
approximation divides $c_n$ into the time-averaged (over $\tau_f$) part
$\bar{c}_n$ and the small (at least as ${\cal O}(\eps)$) oscillating
part: $c_n=\bar{c}_n+\delta c_n$ \cite{bogo}.  To leading order we neglect in the RHS
of (\ref{7}) all the oscillating factors and substitute $c\to\bar{c}$
and $R\to\bar{R}$:
\BEA
\bar{c}^\bullet_n
=-{\sum}_k \bar{c}_k\langle n|k'\rangle\,\overline{F(\R,\bar{c},e^{i\Delta \gamma})\,
e^{i(\gamma_k-\gamma_n)}},
\label{8}
\EEA
where $\tau=\eps t$, $\bar{X}\equiv\int_0^{\tau_f}\frac{\d
s}{\tau_f}\,X(s)$, and where $X^\bullet \equiv {\d X}/{\d \t}$.

{\it Linear feedback}.
The simplest example of feedback is
\BEA
\label{10}
F=\langle \psi |A|\psi\rangle={\sum}_{nm}c_n^*\,c_m \,A_{nm}\,
e^{i (\gamma_m-\gamma_n)},
\EEA
where $\langle n|A|m \rangle\equiv A_{nm}$. Eq.~(\ref{10}) can be regarded as
the first term of the Taylor expansion assuming that $F(x)$
depends weakly on its argument. Eq.~(\ref{8}) leads to
\BEA
\bar{c}_l^\bullet
=-{\sum}_{knm} \bar{c}_k\,\,\langle l| k'\rangle\,\,
\bar{c}_n^*\,\bar{c}_m A_{nm}\,\,
\overline{e^{i (\gamma_m-\gamma_n+\gamma_k-\gamma_l)}}.
\label{12}
\EEA

In working out (\ref{12}) we shall assume that the 
time-integrated energy-level differences are distinct:
\BEA
\label{13}
\gamma_m(t)-\gamma_n(t)\not =\gamma_l(t)-\gamma_k(t), 
\,\,\, {\rm if} \,\, m\not= n \,\,\, {\rm and} \,\, m\not= l.
\EEA
This condition is generic for few-level systems. It
does not hold for cases like harmonic oscillator, which should be 
separately worked out from (\ref{12}).   
Now in the RHS of (\ref{12}) the non-zero terms are those with $m=n$ and
$l=k$, and those with $m=l$ and $k=n$ (but $n\not =l$, not to count
twice the term $m=n=k=l$):
\BEA
\bar{c}_l^\bullet=-\bar{c}_l\,\langle l|l'\rangle\, \R^\bullet
-\bar{c}_l\,{\sum}_{n(\not = l)} |\bar{c}_n|^2\, \langle l|n'\rangle\,A_{nl},
\label{14}
\EEA
where $\langle l| l'\rangle$ is imaginary, since $\partial_R\langle l|l\rangle=1$.
The nontrivial (second) term in the RHS of (\ref{14}) is due to 
non-diagonal elements of $A$. Defining the phase and module of $\bar{c}_n$,
\BEA
\bar{c}_n=\sqrt{p_n}\, e^{i\phi_n },\qquad {\sum}_n p_n=1,
\EEA
we get from (\ref{14}) [and likewise from (\ref{2}, \ref{10})]
\BEA
\label{16}
&&p^\bullet_l=-2p_l{\sum}_{n(\not =l)}p_n\,\Re\{\,
\langle l|n'\rangle\,A_{nl}\,\},\\
\label{17}
&&\phi^\bullet_l=i\langle l|l'\rangle \,R^\bullet-
{\sum}_{n(\not =l)}p_n\,\Im\{\,
\langle l|n'\rangle\,A_{nl}\,\},\\
\label{17.1}
&&\R^\bullet={\sum}_n p_n A_{nn}.
\EEA
Eqs.~(\ref{16}--\ref{17.1}) are our central results. Before exploring
them in more detail let us discuss the standard (open-loop, i.e., no
feedback) adiabatics, where $A=A(R)$ is a c-number. Now $R$ moves in a
prescribed way according to $R^\bullet=A(R)$.  Eq.~(\ref{16}) leads to
the conservation of the probabilities $p^\bullet_l=0$ (adiabatic theorem):
the system does not get enough energy to move out of the given energy level \cite{messiah}.
The RHS of (\ref{17}) reduces to Berry's factor
$\phi^\bullet_{{\rm B},l} =i\langle l|l'\rangle \,R^\bullet$. As seen
from (\ref{5}), though $\phi_{{\rm B},l}$ is by itself not
gauge-invariant, it does bring an observable (Berry phase) contribution
in a non-diagonal average over the state $|\,\psi(t)\,\rangle= \sum_n
c_n(0)e^{i\phi_{{\rm B},n}(\t)+i\gamma_n(t)}$. The Berry phase was observed
in numerous experiments; see \cite{berry,sj_ajp} for review. It
is constrained by the following conditions. 
{\it 1.} The Berry phase nullifies, $\langle l|l'\rangle=0$, if the
adiabatic eigenvectors $|l\rangle$ can be made real via a gauge
transformation, e.g., a spinless particle without magnetic field.
(This statement does not hold if there are level-crossings.)
{\it 2.} $\phi_{{\rm B},l}=0 $ for a cyclic motion of a single slow
parameter $R$, where $R$ is switched on at
the initial time and then switched off at the final time. 
The Berry phase may be
different from zero if there is more than one slow parameter ${\bf
R}=(R_1,R_2,...)$ on a closed curve ${\cal C}$: ${\bf R}(0)={\bf
R}(\tau)$ \cite{berry}. Then one gets a gauge-invariant expression
$\phi_{{\rm B},l}=i\oint_{\cal C} \d {\bf R}\, \langle l|\partial_{{\bf
R}}\, l\rangle$ \cite{berry,sj_ajp}. 

{\it Closed-loop adiabatics.} Eq.~(\ref{16}) for $p_l$ arises out of
the averaging over the fast dynamic phases under condition (\ref{13}).
Eq.~(\ref{16}) is non-linear over $p_n$ due to the feedback. The
probabilities $p_n$ are no longer conserved [due to the resonance between
the oscillations of $c_n$ and those of $R$, see (\ref{12})], and if $p_n$'s
are known, the phases $\phi_l$
are obtained directly from (\ref{17}). The matrix
\BEA
a_{ln}\equiv-
2\Re\{\,\langle l|\partial_R n\rangle\,\langle n|A|l\rangle\,\}, \qquad
a_{ln}=-a_{ln}, 
\label{77}
\EEA
in (\ref{16}) is antisymmetric; in particular, $a_{ll}=0$, which means
$\sum_{l}p_l(\tau)=1$.  The edges of the probability simplex, e.g.
$p_l=\delta_{l1}$, are (possibly unstable) stationary solutions of
(\ref{16}), and $p_l(\t)$ is always non-negative. 

It is noteworthy that (\ref{16}) coincides with the replicator equation
for a zero-sum population game \cite{akin,hofbauer}. Consider a
population of agents that consists of groups $l=1,..,d$. The fraction
$p_l$ of each group in the total population changes due to interaction
between the groups, so that $p_l^\bullet$ is proportional to
$p_l$ itself ({\it autocatalyst principle}), while the proportionality
coefficient is the average payoff of the group $l$:
$p_l^\bullet=p_l\sum_{n}a_{ln}p_n$ \cite{akin,hofbauer}. Here the payoff
matrix $a_{ln}$ determines the gain (or the fitness increase) of the group
$l$ in its interaction with the group $n$. The actual mechanism of this
interaction depends on the concrete implementation of the model
(inheritance, learning, imitation, infection, etc) \cite{hofbauer}.  The
condition $a_{nl}=-a_{ln}$ means a zero-sum game (e.g., poker): the gain
of one group equals to the loss of the other.  Thus in
(\ref{16}) the population game, with (in general)
$\tau$-dependent payoffs $a_{ln}$, is now played by the energy levels.
Interesting features of the replicator equation can be found without
solving it; see (\ref{tamo}--\ref{kaspar}). 

For the open-loop control changing of $R$ on the slow time-scale is
mandatory, otherwise no adiabatic motion occurs at all. The closed-loop
situation is different, since now for $\langle n|A|n\rangle=0$ the slow
motion of $R$ is absent, $\R^\bullet=0$ [see (\ref{17.1})], with still
non-trivial adiabatic dynamics. Thus $R$ does move on the fast
time, but this motion averages out on the slow time. Let us focus on
this situation, since we cannot study (\ref{16}--\ref{17.1}) in full
generality. 

Eqs.~(\ref{16}, \ref{77}), now with $\tau$-independent $a_{ln}$, is conveniently
studied via the time-averages \cite{hofbauer}:
\BEA
\label{tamo}
\frac{1}{T} \ln \frac{p_l(T)}{ p_l(0)}
=\sum_{n}a_{ln}\bar{p}_n(T), \,\,\,
\bar{p}_n(T)=\int_0^T\frac{\d\tau }{T}\, p_n(\tau).
\EEA
There are now two different dynamic scenarios depending on 
the concrete form of $\tau$-independent $a_{lp}$ in
(\ref{16}, \ref{77}).

\begin{figure}
\includegraphics[width=6.5cm]{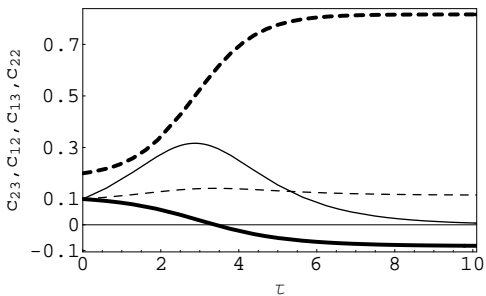}
\caption{Adiabatic amplitudes $c_{nl}$ versus $\tau$ for a three-level system. $c_{nl}$ 
are obtained from solving 
(\ref{36}) in the case $A=-\partial_R H$ (hybrid dynamics). The external field is not acting
on the first energy level: $|\langle 2|1'\rangle|=|\langle 3|1'\rangle|=0$; 
thus $c_{11}$ is constant. We put  $|\langle 2'|3\rangle|=1$. All $|k\rangle$ are real (no Berry phases),
and additionally the adiabatic energies are constant: $A_{ll}=-\partial_R E_l=0$; thus $\bar{R}$
is constant according to (\ref{37}).
Normal curve: $c_{23}$.  Dashed curve: $c_{12}$.
Thick curve: $c_{13}$.
Thick dashed curved: $c_{22}$. Recall that $c_{33}+c_{22}={\rm const}$. 
}
\label{f_1} 
\end{figure}

{\bf 1.} If all $p_l(t)$ (which were non-zero at the initial time
$\tau=0$) stay non-zero for all times, $\ln p_l(T)$ in the LHS of
(\ref{tamo}) is limited, which means that this LHS can
be neglected for $T\to\infty$. We then get from (\ref{tamo})
\cite{akin,hofbauer}
\BEA
\label{kamo}
{\sum}_{n}a_{ln}\bar{p}_n(\infi)
=0.
\EEA
Thus all $p_l(t)$ may remain non-zero for all times provided that there is
a probability vector $\bar{p}(\infi)$ that satisfies (\ref{kamo}).
Clearly, $\bar{p}(\infi)$ is a stationary state of (\ref{16}, \ref{77}).
Recall that the [non-negative] relative entropy is defined as
\BEA
\label{belo}
S[\bar{p}(\infi)|p(\t)]={\sum}_l\bar{p}_l(\infi)\ln\left[\,\bar{p}_l(\infi)/p_l(\t)\,\right],
\EEA
where $p(t)$ is a time-dependent solution of (\ref{16}). $S[\bar{p}_n(\infi)|p(\t)]$ is
equal to zero if and only if $\bar{p}(\infi)=p(\tau)$. Due to (\ref{kamo}), $S[\bar{p}_n(\infty)|p(t)]$ 
is a constant of motion [thus an adiabatic invariant], since 
\BEA
S^\bullet[\bar{p}(\infi)|p(\t)]=
{\sum}_{ln}p_l(\tau)a_{ln}\bar{p}_n(\infi ).
\label{kaspar}
\EEA 
This constant can be associated with
Hamiltonian, and then (\ref{16}) can be recast into a Hamiltonian
form \cite{akin}. The non-linearity of this dynamics is
essential, since it can demonstrate chaos for $d\geq 5$
\cite{aa}.  In some closely related systems the chaotic behavior 
was seen in \cite{pico}.

{\bf 2.} If the matrix $a_{ln}$ is such that (\ref{kamo}) does not have
any probability vector solution, $\frac{1}{T}\ln \frac{p_l(T)}{p_l(0)}$ in
(\ref{tamo}) is necessarily
finite for at least one $l$. The corresponding probability $p_l(T)$ goes
to zero (for a large $T$): $p_l(T)\to p_l(\infi)=0$, so that for all $k$
one has ${\sum}_{n}a_{kn}\bar{p}_n(\infi)\leq 0$. 
This inequality is strict at least for $k=l$. Eq.~(\ref{kaspar}) 
shows that $S[\bar{p}(\infi)|p(\t)]$ now decays to zero
meaning that $p(\t)$ {\it relaxes} to $\bar{p}(\infi)$.  This relaxation
is due to the non-linearity of (\ref{16}); it is impossible without
feedback. 

Eq.~(\ref{17}) for the phases integrates as
\BEA
\label{metro}
\phi_l(\tau)=-\tau {\sum}_{n(\not =l)}\bar{p}_n(\tau)b_{ln}, \,
b_{ln}\equiv \Im\{\langle l|n'\rangle\,\langle n|A|l\rangle\},
\EEA
where $\bar{p}_n(\tau)$ satisfies to the algebraic equation
(\ref{tamo}).  Note that $b_{ln}$ is symmetric: $b_{ln}=b_{nl}$.
Eq.~(\ref{metro}) gives the phases of the adiabatic (linear) feedback
control.  It is clear that $\phi_l(\tau)$ is free of the constraints for
the open-loop (Berry) phase $\phi_{{\rm B},l}$: {\it i)} it is
explicitly gauge-invariant together with $b_{ln}$; {\it ii)} its
existence does not require complex adiabatic eigenvectors $|l\rangle$,
provided that the monitoring observable $A$ has at least some complex
elements $\langle n|A|l\rangle$; {\it iii)} it does not require several
control parameters for cyclic processes; {\it iv)} even if $a_{nl}$,
defined via (\ref{77}), is zero, i.e., if the probabilities $p_n$ are
conserved, the feedback-driven phases $\phi_l$ in (\ref{metro}) can be
non-zero. Note that $\phi_l=0$ if the evolution starts from a strictly one
adiabatic eigenvector $p_n(0)=\delta_{nk}$ (however this stationary state of
(\ref{16}) need not be stable, as we saw above). 


{\it Examples.} We now apply our findings to two simple examples.
For a two-level system (\ref{16}, \ref{77}) reduce to
$p_1(\t)=\frac{p_1(0)e^{a_{12}\t}}{1+p_1(0)[e^{a_{12}\t}-1]}$,   
which means that independent of the initial value $p_1(0)$, $p_1\to
1$ ($p_1\to 0$) if $a_{12}>0$ ($a_{12}<0$).  Properly choosing the time
$\t$ and $a_{12}$, and knowing $p_1(0)$, we can reach any value $0\leq
p_1(\tau)\leq 1$. Eq.~(\ref{metro}) leads to
$\phi_{1,2}(\tau)=\pm\frac{b_{12}}{a_{12}}\ln\left[
p_1(0)(e^{\mp a_{12}\tau}-1)+1\right]$.
Two basic examples of two-level systems are the spin-$\frac{1}{2}$ and
the polarization states of light. The standard Berry phase was observed 
in both these cases \cite{berry}.

For the three-level situation the internal stationary vector is obtained from (\ref{kamo})
(up to normalization)
\BEA
\label{kogan_1}
\bar{p}_1(\infi)\propto   {a_{23}}/{a_{12}}, \quad 
\bar{p}_2(\infi)\propto - {a_{13}}/{a_{12}}, \quad 
\bar{p}_3(\infi)\propto 1,
\EEA
provided these probabilities are positive, i.e., (using game-theoretic
terms) the group $1$ beats $2$, $2$ beats $3$, but $3$ beats $2$
(rock-swissor-paper game).  Now the $\tau$-dependent solution $p(\tau)$
of (\ref{16}) oscillates around (\ref{kogan_1}). Naturally, if one group
(say $1$) beats both others ($a_{12}>0$, $a_{13}>0$), the only attractor
of (\ref{16}) is $\bar{p}(\infi)=(1,0,0)$. The latter conclusion holds
also for a $\tau$-dependent $\bar{R}$, if the conditions
$a_{12}(\tau)>0$ and $a_{13}(\tau)>0$ are satisfied for all $\tau$'s.
However, the general arguments (\ref{tamo}--\ref{kaspar}) do not
proceed for $\tau$-dependent $a_{ln}$. 

{\it Mixed states. } So far we focussed on pure states of the quantum
system.  Now we assume that the quantum state $\rho$ is mixed and the
feedback goes via the average ${\rm tr}(A\rho)$; compare with (\ref{2}).
Since the closed loop equations (\ref{1}) is not linear, the mixed-state dynamics
(in general) does not reduce to the pure case.  Defining for the
adiabatic amplitude $c_{nm}\equiv \langle n|\rho|m\rangle
e^{i\gamma_m-i\gamma_n}$ [compare with (\ref{5})], and proceeding along the lines of
(\ref{3}--\ref{13}) leads to
\BEA
\label{36}
&&\mc_{nm}^\bullet+\R^\bullet \mc_{nm}(\langle n|n'\rangle+\langle m'|m\rangle)\\
&&=-{\sum}_{l(\not =n)} \langle n|l'\rangle A_{ln}\mc_{nl}\mc_{lm}
- {\sum}_{l(\not =m)} \langle l'|m\rangle \mc_{nl}\mc_{lm}A_{ml}\nonumber\\
\label{37}
&&\R^\bullet={\sum}_l \mc_{ll}A_{ll}.
\EEA
There is a case where the pure-state analysis applies directly:
Pseudo-pure states in NMR are important for ensemble computation and are
given as $\rho=(1-\eta)\frac{\hat{1}}{d}+\eta |\psi\rangle\langle\psi|$,
where $\hat{1}$ is the unit matrix, and where $0<\eta<1$ is a parameter
\cite{jajones}.  Since $\hat{1}$ is an invariant of (\ref{36}),
Eq.~(\ref{36}) reduces to (\ref{14}), but with $A_{nl} \to\eta^2 A_{nl}$.
In general the phases of $c_{nm}$ do not decouple from $|c_{nm}|$,
and we do not have a general theory for mixed states. 

Let us study in more detail
the hybrid dynamics $A=-\partial_R H$; see our
discussion after (\ref{2}). This is a pertinent example, since its pure-state dynamics is
straightforward: due to $A_{ln}=(E_n-E_l)|\langle l'|m\rangle|^2$, 
the new phases (\ref{metro}) nullify, while (\ref{16}) for the probabilities predicts relaxation
to the ground state (cooling). The mixed-state dynamics will be seen to be more interesting.
Eq.~(\ref{36}) implies 
\BEA
\mc_{nn}^\bullet
=2{\sum}_{l(\not =m)} (E_l-E_n)|\langle l'|m\rangle|^2 |\mc_{nl}|^2.
\EEA
Let $n=1$ be the lowest energy level. If all $|\langle l'|1\rangle|$
differ from zero, $c_{l\not =1}$ has to nullify for large $\tau$, since
$c_{11}$ should be limited.  Continuing this way for $n>1$, we get
that all non-diagonal elements $c_{n\not = l}$ nullify (decoherence), if
all $|\langle l'|n\rangle|$ are positive.  

If, however, $|\langle l'|n\rangle|=0$ for some $n\not
=l$, the element $c_{nl}$ survives and undergoes a
non-trivial evolution.  An example of this is presented in
Fig.~\ref{f_1}. Here the field $R$ acts only on energy levels $2$ and
$3$; the level $1$ does not feel $R$.  Thus $|\langle
2|1'\rangle|=|\langle 3|1'\rangle|=0$ and $|\langle 2'|3\rangle|>0$.
Nevertheless, $c_{12}$ and $c_{13}$ do change in time. In Fig.~\ref{f_1}
all $c_{l\not = n}$ are real, and we see that $c_{13}$ changes its sign,
an example of the adiabatic phase. Though $c_{23}$ has to decay to zero,
it can increase in the intermediate times.

{\it In summary}, we studied how the feedback generated by
non-disturbing (ensemble) measurements affects the adiabatic (i.e.,
slowly driven) quantum dynamics. For the simplest
linear feedback we have found that {\it i)} the populations are no
longer constant. Instead, they satisfy the canonical [replicator]
equation of the population game theory, allowing us to visualize the
corresponding dynamics as a zero-sum game played by the adiabatic energy
levels. The [non-linear] replicator equation generates a non-trivial
(possibly chaotic) Hamiltonian motion, or alternatively, relaxation towards
a certain state.  {\it ii)} In addition to the
Berry phase, the feedback generates a new, explicitly gauge-invariant
phase, which [as compared to the Berry phase] exists under a wider range
of conditions.  In particular, there are scenarios of feedback, where
the probabilities are constant (resembling the ordinary situation), but
the new phases are still non-trivial. These results extend to 
pseudo-pure quantum states.

The work was supported by Volkswagenstiftung grant
``Quantum Thermodynamics: Energy and information flow at nanoscale''.

\end{document}